\documentclass[12pt,a4paper,twoside]{article}

\usepackage{amsmath}
\usepackage{amssymb}
\usepackage{theorem}
\usepackage{subdepth}
\usepackage{graphicx}
\usepackage{tikz}
\usetikzlibrary{positioning}

\newcommand{\N}{\mathbb N}
\newcommand{\Z}{\mathbb Z}
\newcommand{\R}{\mathbb R}
\newcommand{\C}{\mathbb C}
\newcommand{\e}{{\rm e}}
\newcommand{\pf}{{\rm pf}}
\renewcommand{\Re}{{\rm Re}}
\renewcommand{\Im}{{\rm Im}}
\renewcommand{\i}{{\rm i}}
\def\slim{\mathop{\rm s-lim}}

\newcommand{\Aut}{{\rm Aut}}
\newcommand{\sign}{{\rm sign}}
\newcommand{\spec}{{\rm spec}}
\newcommand{\eig}{{\rm eig}}
\newcommand{\pp}{{\rm pp}}
\newcommand{\ac}{{\rm ac}}
\renewcommand{\sc}{{\rm sc}}
\renewcommand{\d}{{\rm d}}
\newcommand{\fh}{\mathfrak h}
\newcommand{\fH}{\mathfrak H}
\newcommand{\fA}{{\mathfrak A}}
\newcommand{\ff}{{{\mathfrak f}}}
\newcommand{\fF}{{\mathfrak F}}
\newcommand{\mE}{\mathcal E}
\newcommand{\mH}{\mathcal H}
\newcommand{\mL}{\mathcal L}

\newcommand{\tr}{{\rm tr}}

\def\bas#1\eas{\begin{align*}#1\end{align*}}
\def\ba#1\ea{\begin{align}#1\end{align}}
\newcommand{\br}{\begin{remark}}
\newcommand{\er}{\end{remark}}
\newcommand{\bd}{\begin{definition}}
\newcommand{\ed}{\end{definition}}
\newcommand{\bt}{\begin{thm}}
\newcommand{\et}{\end{thm}}
\newcommand{\bprf}{\noindent{\it Proof.}}
\newcommand{\eprf}{\hfill $\Box$}
\newcommand{\bp}{\begin{proposition}}
\newcommand{\ep}{\end{proposition}}
\newcommand{\bc}{\begin{corollary}}
\newcommand{\ec}{\end{corollary}}
\newcommand{\intg}{\int_{-\pi}^\pi\frac{{\rm d}k}{2\pi}\hspace{0.5mm}}

\newcommand{\rd}{\hspace{-0.5mm}{\rm d}}

\newtheorem{thm}{Theorem}
\newtheorem{definition}[thm]{Definition}
{\theorembodyfont{\upshape} \newtheorem{remark}[thm]{\it Remark}}
\newtheorem{proposition}[thm]{Proposition}
\newtheorem{corollary}[thm]{Corollary}

\begin{document}
\pagestyle{myheadings}
\markboth{Walter H. Aschbacher}{A rigorous scattering approach to quasifree
fermionic systems out of equilibrium}

\title{A rigorous scattering approach to quasifree fermionic systems out of 
equilibrium}

\author{Walter H. Aschbacher\footnote{walter.aschbacher@univ-tln.fr}\\
\\
Universit\'e de Toulon, Aix Marseille Univ, CNRS, CPT, Toulon, France}

\date{}
\maketitle

\begin{abstract}
Within the rigorous axiomatic framework for the description of quantum mechanical
systems with a large number of degrees of freedom, we construct the so-called nonequilibrium steady state for the quasifree fermionic system corresponding to the 
isotropic XY chain in which a finite sample, subject to a local gauge breaking 
anisotropy  perturbation, is coupled to two thermal reservoirs at different 
temperatures.  Using time dependent and stationary scattering theory, we 
rigorously prove, from first principles, that the nonequilibrium system under consideration is thermodynamically nontrivial, i.e., that its entropy production rate is strictly positive.
\end{abstract}


\noindent {\it PACS (2010)}\, 
02.30.Tb, 03.65.Nk, 71.10.Ca, 72.10.Fk, 65.40.gd\\


\section{Introduction}

In recent years, a wide range of important thermodynamic properties of open
quantum systems have successfully been derived from first principles. A precise 
analysis of such  systems having a large, i.e., often, in physically idealized 
terms, an infinite number of degrees of freedom, is most clearly carried out 
within the axiomatic framework of operator algebras. As a matter of fact, after 
having been heavily used in the 1960s, in particular for the description of 
quantum systems in thermal equilibrium (see, for example, \cite{BR}), the 
benefits of this framework have again started to unfold more recently in the 
physically much more general situation of open quantum systems out of 
equilibrium. In the latter field, most of the rare mathematically rigorous results 
have been obtained for the so-called nonequilibrium steady states (NESSs) 
introduced in \cite{R2001} by means of scattering theory on the algebra of 
observables. 

In quantum statistical mechanics both in and out of equilibrium, an important 
role is played by the so-called quasifree fermionic systems, and this is true 
not only because of their mathematical accessibility but also when it comes to 
real physical applications. Indeed, from a mathematical point of view, these 
systems allow for a simple and powerful description by means of scattering 
theory restricted to the underlying 1-particle Hilbert space over which the 
fermionic algebra of observables is constructed. The restriction of the dynamics 
to the 1-particle sector opens the way for a rigorous mathematical analysis of
many properties which are of fundamental physical interest. But, beyond their
importance due to their mathematical accessibility, quasifree fermions also 
constitute a class of systems which effectively describe nature. Aside from the 
various electronic systems in their independent electron approximation, they also 
play a part in the rigorous approach to spin systems. One of the most prominent
representatives of the latter is the so-called XY spin chain, introduced in 1961 in
 \cite{LSM1961}, for which a physical realization has already been identified in 
the late 1960s (see, for example, \cite{CSP1969}). Its impact on the interplay between 
the experimental, numerical, theoretical, and mathematical research activity in 
the field of low-dimensional magnetic systems is being felt ever since (see, for 
example, \cite{MK2004}).

\vspace{1mm}
In the present paper, we rigorously analyze, from first principles, the entropy 
production rate in the quasifree fermionic system over the two-sided discrete 
line $\Z$ which, in the spin picture, corresponds to the isotropic XY spin chain 
perturbed by a local anisotropy (in contrast to \cite{AP2003}, where the anisotropy 
acts as a global and homogeneous perturbation).

In order to specify the desired nonequilibrium configuration, we first fix 
$n\in\Z$ with $n\ge 1$ and cut the finite piece 
\ba
\Z_S
&:=\{x\in\Z\,|\, |x|\le n\}
\ea
of length $2n+1$ out of the two-sided discrete line. This piece plays the role 
of the configuration space of the confined sample, whereas the remaining  infinite 
parts,
\ba
\Z_L
&:=\{x\in\Z\,|\, x\le -(n+1)\},\\
\Z_R
&:=\{x\in\Z\,|\, x\ge n+1\},
\ea
act as the configuration spaces of the extended thermal reservoirs. Over these
configuration spaces, we define the initial state to be the decoupled product of 
three thermal equilibrium states carrying the corresponding inverse temperatures
\ba
\label{betas}
0=
\beta_S
<\beta_L
<\beta_R
<\infty.
\ea
The NESS is then constructed with respect to the full time evolution which, by 
definition, not only couples the sample to the reservoirs but also exposes the
sample to a local anisotropy perturbation of strength 
\ba
\gamma\in\R,
\ea

whose support resides on the sites $\{a, a+1\}$, where $a\in\Z$ satisfies 
\ba
\label{a}
-n
\le a
\le  n-1.
\ea

\vspace{5mm}
The paper is organized as follows.

\vspace{1mm}
{\it Section \ref{sec:setting}}\, specifies the nonequilibrium setting we are 
interested in, i.e., it introduces the canonical anticommutation relation (CAR) 
algebra of observables, its selfdual version, the quasifree dynamics generated by
the 1-particle Hamiltonians, and the quasifree initial state. 

\vspace{0.7mm}
{\it Section \ref{sec:ness}}\, is devoted to the definition and the construction of 
the NESS in the nonequilibrium setting at hand. It turns out that the absolutely
continuous part of its 2-point operator, computed by means of time dependent 
scattering theory, is determined through the so-called intermediate wave operator 
which, by definition, compares the free isotropic XY dynamics with the free isotropic 
XY dynamics perturbed by a local anisotropy. 

\vspace{0.7mm}
{\it Section \ref{sec:scattering}}\, contains the derivation of the action of the
intermediate wave operator using stationary scattering theory. Due to the fact that 
the local anisotropy is a 2-site perturbation breaking gauge invariance, the action 
of the intermediate wave operator is substantially more complicated than the one 
from \cite{A2011} for a gauge invariant local 1-site perturbation.

\vspace{0.7mm}
{\it Section \ref{sec:entropy}}\, introduces the notions of heat flux and entropy
production rate. A general formula is derived for the NESS expectation value of
the extensive energy current observable describing the energy flow through the 
sample as a function of the strength of the anisotropy perturbation. It is proven 
that the nonequilibrium system under consideration is thermodynamically nontrivial, 
i.e., that its entropy production rate, the first fundamental physical quantity for 
systems out of equilibrium, is strictly positive.

\section{Nonequilibrium setting}
\label{sec:setting}

In this section, we specify the nonequilibrium setting we are interested in. 
First recall that, in the operator algebraic formalism of quantum statistical 
mechanics, a physical system is characterized by an algebra of observables, by
a group of time evolution automorphisms, and by a normalized positive linear 
state functional on the observable algebra (see, for example, \cite{BR} for a 
detailed description of this formalism). Definitions \ref{def:obs}, \ref{def:dyn}, 
and \ref{def:initial} below spell out the corresponding three ingredients for 
the quasifree nonequilibrium setting to be studied. Here and there, we will also 
make brief remarks on the underlying general framework.

\vspace{1mm}
In the following, for all complex Hilbert spaces $\mH$, we denote by $\mL(\mH)$ 
and $\bar \mL(\mH)$ the sets of bounded linear and antilinear operators on $\mH$,
respectively. Moreover, $\mL^0(\mH)$ stands for the finite rank operators and 
$\mL^1(\mH)$ for the trace class operators on $\mH$. For elements $A,B$ in the 
various sets in question below, the commutator and the anticommutator of  $A$ 
and $B$ are denoted by $[A,B]:=AB-BA$ and $\{A,B\}:=AB+BA$, respectively. 
Finally, $\sigma_1, \sigma_2, \sigma_3\in\C^{2\times 2}$ are the usual Pauli matrices 
and, for all $m\in\N$, we denote by $\C^{m\times m}$ the complex $m\times m$
matrices.

\bd[Observables]
\label{def:obs}
\hspace{0mm}

\vspace{1mm}
\noindent{\it (a)} 1-particle Hilbert space\\
Let $\Z$ be the configuration space of the system and let 
\ba
\fh
:=\ell^2(\Z)
\ea
be the separable complex 1-particle Hilbert space of square-summable complex-valued
functions on $\Z$. Moreover, we set $\fH:=\fh\oplus\fh$ and, on both $\fh$ and $\fH$, 
we denote the scalar products and the corresponding induced norms by $(\cdot,\cdot)$
and $\|\cdot\|$, respectively.

\vspace{1mm}
\noindent{\it (b)} Algebra of observables\\
The algebra of observables is defined to be the CAR algebra over $\fh$, denoted by
\ba
\fA
:={\rm CAR}(\fh),
\ea
whose generators are written, as usual, as $1$, $a(f)$, and $a^\ast(f)$ for all $f\in\fh$.

\vspace{1mm}
\noindent{\it (c)} Selfdual generators\\
The complex linear map $B:\fH\to\fA$, defined, for all 
$F:=f_1\oplus f_2\in\fH$, by
\ba
B(F)
:=a^\ast(f_1)+a(\zeta f_2),
\ea
where $\zeta f:=\bar f$ stands for the complex conjugation on $\fh$, satisfies the relations 
\ba
\label{B1}
B^\ast(F)
&=B(\Gamma F),\\
\label{B2} 
\{B^\ast(F), B(G)\}
&=(F,G) 1,
\ea
where the antiunitary involution $\Gamma\in\bar\mL(\fH)$ is defined to act as the
operator matrix $\Gamma:=\zeta\sigma_1$ on the direct sum $\fH=\fh\oplus\fh$, and
\eqref{B2}, called the selfdual CARs, follows from the usual CARs. 

\vspace{1mm}
\noindent{\it (d)} Selfdual second quantization\\
The complex linear map $b:\mL^0(\fH)\to\fA$, defined, for all $m\in\N$, all 
$F_1,\ldots,F_m\in\fH$, all $G_1,\ldots, G_m\in\fH$, and 
$A:=\sum_{i=1}^m(F_i,\hspace{0.5mm}\cdot\hspace{0.5mm})G_i$, by
\ba
\label{sdsq}
b(A)
:=\sum_{i=1}^m B(G_i)B^\ast(F_i),
\ea
is called the selfdual second quantization of $A$.
\ed

\br
The algebra of observables $\fA$ is a so-called $C^\ast$-algebra. It is 
$\ast$-isomorphic to $\overline{\rm SDC}(\fH,\Gamma)$, the ($C^\ast$-completed) 
selfdual CAR algebra over $\fH$ and $\Gamma$. The selfdual framework is a useful
general concept which has been developed in \cite{Araki1971} and \cite{Araki1987}
 (see there for a more detailed description of the selfdual objects used in the 
following).
\er

\br
Note that \eqref{sdsq} does not depend on the choice of the functions 
$F_1,\ldots,G_m$ which represent $A$. Moreover, the definition of $b$ can be 
extended to $\mL^1(\fH)$ using the fact that $\mL^0(\fH)$ is dense in $\mL^1(\fH)$ 
with respect to the so-called trace norm.
\er

We next specify the 2nd ingredient. As discussed in the introduction, the Hamiltonians
which we will introduce describe the decoupling of the reservoirs from the sample and
the coupled system with and without the local anisotropy perturbation.

\vspace{1mm}
In the following, the completely localized elements of the orthonormal Kronecker
basis $\{\delta_x\}_{x\in\Z}$ of $\fh$ are given, for all $x,y\in\Z$, by $\delta_x(y):=1$ 
if $y=x$ and $\delta_x(y):=0$ if $y\neq x$. Moreover, for all $A\in\mL(\mH)$, we will 
use the notation $\Re[A]:=(A+A^\ast)/2$ and $\Im[A]:=(A-A^\ast)/(2\i)$. 

\bd[Dynamics]
\label{def:dyn}
\hspace{0mm}

\vspace{1mm}
\noindent{\it (a)} 1-particle Hamiltonians\\
Using the right translation $u\in\mL(\fh)$ and the localization operator 
$p_{x,y}\in\mL(\fh)$, given by $(uf)(x):=f(x-1)$ and $p_{x,y}f:=f(x)\delta_y$ for all
$f\in\fh$ and all $x,y\in\Z$, we define
\ba
h
&:=\Re[u],\\
\label{vd}
v_\d
&:=\Re[u^{-n} p_{0,0} u^{n+1}]+\Re[u^{n+1} p_{0,0} u^{-n}],\\
\label{v}
v
&:=\Im[p_{a+1,a}].
\ea
The liftings to $\mL(\fH)$ are given by $H:=h\hspace{0.1mm}\sigma_3$,
$V_\d:=v_\d\hspace{0.1mm}\sigma_3$, $H_\d:=H-V_\d$, and 
\ba
\label{V}
V
&:=v\hspace{0.2mm}\sigma_2,\\
\label{Hgam}
H_\gamma
&:=H+\gamma V.
\ea
The Hamiltonians $H$ and $H_\d$, diagonal with respect to $\fH=\fh\oplus\fh$, are
called the XY Hamiltonian and the decoupled Hamiltonian, respectively, whereas
$H_\gamma$ is non-diagonal and called the anisotropy Hamiltonian.

\vspace{1mm}
\noindent{\it (b)} Dynamics\\
The quasifree dynamics generated by the XY Hamiltonian, the decoupled Hamiltonian,
and the anisotropy Hamiltonian are defined, for all $t\in\R$ and all $F\in\fH$, by 
\ba
\label{tau}
\tau^t(B(F))
&:=B(\e^{\i t H}F),\\
\label{tau0}
\tau_\d^t(B(F))
&:=B(\e^{\i t H_\d}F),\\
\label{taug}
\tau^t_\gamma(B(F))
&:=B(\e^{\i t H_\gamma}F),
\ea
and by a suitable extension to the whole of $\fA$ (see Remark \ref{rem:ham}). The
dynamics $\tau^t$, $\tau_\d^t$, and $\tau^t_\gamma$ are called the XY dynamics,
the decoupled dynamics, and the anisotropy dynamics, respectively.
\ed

\br
\label{rem:ham}
In the selfdual framework, an operator $A\in\mL(\fH)$ is called a Hamiltonian if
$A^\ast=A$ and $\Gamma A\Gamma=-A$. The second condition is a consequence of
\eqref{B1}, of the fact that, by definition, the dynamics $\sigma^t$ generated by $A$ 
as in \eqref{tau} -- \eqref{taug} is, for all $t\in\R$, a $\ast$-automorphism on $\fA$,
i.e., a map from $\fA$ to $\fA$ preserving the vector space structure, the algebra
multiplication, and the $\ast$-operation on $\fA$ (we will denote by $\Aut(\fA)$ the 
set of all such maps), and of $\|F\|/\sqrt{2}\le \|B(F)\|\le \|F\|$ (where $\|B(F)\|$is the
$C^\ast$-norm of $B(F)\in\fA$). Both conditions are satisfied for all the Hamiltonians 
of Definition \ref{def:dyn}{\it (b)}.

In addition, the dynamics $\{\sigma^t\}_{t\in\R}\subseteq\Aut(\fA)$ is defined to be 
a strongly continuous group, i.e., the map $\R\ni t\mapsto\sigma^t\in\Aut(\fA)$ is a 
group homomorphism and, for all $A\in\fA$, the map $\R\ni t\mapsto\sigma^t(A)\in\fA$
is continuous with respect to the $C^\ast$-norm on $\fA$. Such a pair $(\fA,\sigma^t)$ 
is sometimes called a $C^\ast$\hspace{-0.5mm}-dynamical system.
\er

\br
\label{rem:sym}
Defining parity $\theta\in\mL(\fh)$ and the local gauge transformation 
$\xi\in\mL(\fh)$ by $(\theta f)(x):=f(-x)$ and $(\xi f)(x):=\e^{\i \pi x}f(x)$ for all 
$f\in\fh$ and all $x\in\Z$, we get the symmetries
\ba
\label{hu}
[h,u]
&=0,\\
\label{htheta}
[h,\theta]
&=0,\\
\label{hxi}
\{h,\xi\}
&=0,
\ea
which we will use below in the proof of Proposition \ref{prop:wave}.
\er

\br
\label{rem:xy}
The model specified by Definition \ref{def:dyn} has its origin in the XY model whose
Hamiltonian density has the form
\ba
\label{XYDensity}
(1+\gamma)\,\sigma_1^{(x)}
\sigma_1^{(x+1)}+(1-\gamma)\,\sigma_2^{(x)}\sigma_2^{(x+1)},
\ea
where the superscripts denote the sites in $\Z$ of the local Hilbert space of the 
spin chain on which the Pauli matrices act. Indeed, using the so-called 
Araki-Jordan-Wigner transformation introduced in \cite{Araki1984} for 1-dimensional
systems whose configuration space extends infinitely in both directions, 
\eqref{XYDensity}, in the fermionic picture, reads (up to a global prefactor)
\ba
\label{FDensity}
a_x^\ast a_{x+1} + a_{x+1}^\ast a_x
+\gamma (a_x^\ast a_{x+1}^\ast +a_{x+1}a_x),
\ea
where we set $a_x:=a(\delta_x)$ and $a_x^\ast:=a^\ast(\delta_x)$ for all $x\in\Z$.

In order to treat the anisotropic case $\gamma\neq 0$, i.e., the case in which there 
is an asymmetry between the 1st and the 2nd term in \eqref{XYDensity}, the selfdual
quasifree setting is most natural since gauge invariance is broken in \eqref{FDensity}.
Hence, due to the presence of the $\gamma$-term, the anisotropy Hamiltonian 
acquires non-diagonal components with respect to $\fH=\fh\oplus\fh$ (see \eqref{V} 
and \eqref{Hgam}). In many respects, the truly anisotropic XY model is substantially 
more complicated than the isotropic one.
\er

We now arrive at the specification of the initial state, the 3rd and last ingredient 
needed for the construction of the NESS we are interested in. As discussed in the
introduction, it describes the initial configuration in which the left and right reservoirs 
are decoupled from the sample.

\vspace{1mm}
In the following, for all $M\subseteq\R$, we denote by $1_M$ the usual characteristic
function of $M$ on $\R$, i.e., $1_M(x)$ equals $1$ if $x\in M$ and is $0$ otherwise.
Moreover, recall again that a state $\omega$ is a normalized positive linear functional 
on the observable algebra $\fA$, and let us denote by $\mE_\fA$ the set of all states.

\bd[Initial state]
\label{def:initial}
\hspace{0mm}

\vspace{1mm}
\noindent{\it (a)} Fermi-Dirac function\\
For all $\beta\in\R$ and all $e\in\R$, the Fermi-Dirac function is defined by
\ba
\rho_\beta(e)
:=\frac{1}{1+\e^{\beta e}}.
\ea

\vspace{1mm}
\noindent{\it (b)} Projected 1-particle Hamiltonians\\
With the help of the orthogonal projections $p_L, p_R\in\mL(\fh)$, given by
$p_Lf:=1_{\Z_L}f$ and $p_Rf:=1_{\Z_R}f$ for all $f\in\fh$, the 1-particle Hamiltonians
$h_L, h_R\in\mL(\fh)$ are defined by
\ba
h_L
&:=p_Lhp_L,\\
h_R
&:=p_Rhp_R,
\ea
and their liftings to $\mL(\fH)$ by $H_L:=h_L\sigma_3$ and $H_R:=h_R\sigma_3$.

\vspace{1mm}
\noindent{\it (c)} Initial state\\
We define the initial state $\omega_\d\in\mE_\fA$ to be the 
decoupled quasifree state whose 2-point operator $S_\d\in\mL(\fH)$ has the form
\ba
S_\d
&:=(1-s_\d)\oplus \zeta s_\d\zeta,\\
s_\d
&:=\rho_1(\beta_Lh_L+\beta_Rh_R),
\ea
where the operator $s_\d\in\mL(\fh)$ is defined with the help of the spectral 
theorem. 
\ed

\br
\label{rem:qfs}
In the selfdual framework, an operator $S\in\mL(\fH)$ is called a 2-point operator 
if $S^\ast=S$, $0\le S\le 1$, and $\Gamma S\Gamma=1-S$ (the last condition comes, 
in particular, from \eqref{B1}, \eqref{B2}, and \eqref{2ptf} below). For any (not
necessarily quasifree) state $\omega\in\mE_\fA$, there exists a unique 2-point 
operator $S$ such that, for all $F,G\in\fH$, it holds that
\ba
\label{2ptf}
\omega(B^\ast(F)B(G))
=(F,SG).
\ea
If $\omega$ is a quasifree state induced by the 2-point operator $S$, it is completely
characterized by its 2-point function \eqref{2ptf} since, by definition, $\omega$ is 
even and the nonvanishing many-point functions factorize in Pfaffian form, i.e., for 
all $m\in\N$ and all $F_1,\ldots, F_{2m}\in\fH$, we have
\ba
\label{qfs}
\omega(B(F_1)\ldots B(F_{2m}))
=\pf\big([(\Gamma  F_i, S F_j)]_{i,j=1}^{2m}\big),
\ea
and we recall that the Pfaffian is defined by 
$\pf(A):=\sum_{\pi}\sign(\pi) \prod_{i=1}^{m} A_{\pi(2i-1), \pi(2i)}$ for all
$A\in\C^{2m\times 2m}$, where the sum is running over all the $(2m)!/(2^{m}m!)$
pairings of the set $\{1,\ldots, 2m\}$, i.e., over all the permutations $\pi$ in the
permutation group of $2m$ elements satisfying $\pi(2i-1)<\pi(2i+1)$ for all 
$i\in\{1,\ldots, m-1\}$ and $\pi(2i-1)<\pi(2i)$ for all $i\in\{1,\ldots, m\}$ (see Figure \ref{fig:pairings}).

\begin{center}
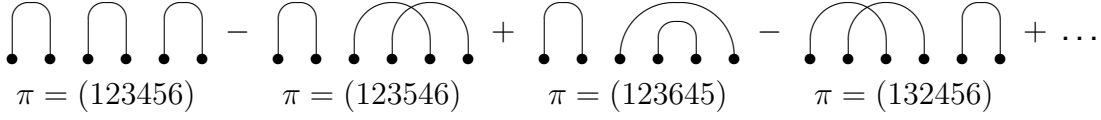
\begin{figure}
\setlength{\unitlength}{1cm}
\begin{center}
\begin{picture}(22,2)
\multiput(1,0)(0.5,0){6}{\circle*{0.15}}
\put(1.25,0){\oval(0.5,1.5)[t]}
\put(2.25,0){\oval(0.5,1.5)[t]}
\put(3.25,0){\oval(0.5,1.5)[t]}
\put(1.05,-0.6){$\pi=(123456)$}
\put(3.8,0.25){$-$}
\multiput(4.5,0)(0.5,0){6}{\circle*{0.15}}
\put(4.75,0){\oval(0.5,1.5)[t]}
\put(6,0){\oval(1,1.5)[t]}
\put(6.5,0){\oval(1,1.5)[t]}
\put(4.55,-0.6){$\pi=(123546)$}
\put(7.3,0.25){$+$}
\multiput(8,0)(0.5,0){6}{\circle*{0.15}}
\put(8.25,0){\oval(0.5,1.5)[t]}
\put(9.75,0){\oval(1.5,1.5)[t]}
\put(9.75,0){\oval(0.5,1.0)[t]}
\put(8.05,-0.6){$\pi=(123645)$}
\put(10.8,0.25){$-$}
\multiput(11.5,0)(0.5,0){6}{\circle*{0.15}}
\put(12,0){\oval(1,1.5)[t]}
\put(12.5,0){\oval(1,1.5)[t]}
\put(13.75,0){\oval(0.5,1.5)[t]}
\put(11.55,-0.6){$\pi=(132456)$}
\put(14.3,0.25){$+$}
\put(14.8,0.25){\ldots}
\end{picture}
\end{center}
\vspace{0.5cm}
\caption{\label{fig:pairings} Some of the pairings for $m=3$. The number of 
intersections $I$ per graph relates to the signature of the corresponding 
permutation $\pi$ as $\sign(\pi)=(-1)^I$.}
\end{figure}
\end{center}
\er

\section{Nonequilibrium steady state}
\label{sec:ness}

In this section, we give a precise definition of the NESS discussed in the introduction.

\bd[NESS]
\label{def:ness}
The state $\omega_\gamma\in\mE_\fA$, defined, for all $A\in\fA$, by
\ba
\label{ness}
\omega_\gamma(A)
:=\lim_{T\to\infty}\frac1T\int_0^T\rd t\hspace{1.5mm}
\omega_\d(\tau^t_\gamma(A)),
\ea
is called the anisotropy NESS associated with the initial state $\omega_\d$ and the
anisotropy dynamics $\tau^t_\gamma$. Moreover, its 2-point operator is denoted by
$S_\gamma\in\mL(\fH)$.
\ed

\br
The general definition, which we already specialized to our case in Definition
\ref{def:ness} (we will see below that the limit in \eqref{ness} exists), stems from
\cite{R2001} and defines the NESSs to be the limit points in the weak-$\ast$ topology 
of the net defined by the ergodic mean between $0$ and $T>0$ of the given initial 
state time evolved by the perturbed dynamics of interest (note that, due to the 
Banach-Alaoglu theorem, the set of such NESSs is not empty). In general, the averaging
procedure allows to treat a nonvanishing contribution to the point spectrum of the
Hamiltonian which generates the full time evolution.
\er

In the following, for all selfadjoint operators $A\in\mL(\fH)$, we denote by 
$1_{\rm ac}(A)$, $1_{\rm sc}(A)$, and $1_{\rm pp}(A)$ the orthogonal projections 
onto the absolutely continuous subspace, the singular continuous subspace, and the 
pure point subspace of $A$, respectively. Moreover, $\eig(A)$ stands for the set of
eigenvalues of $A$, and $1_e(A)$ denotes the spectral projection onto the eigenspace
associated with the eigenvalue $e\in\eig(A)$. Furthermore, the limit with respect to 
the strong operator topology on $\mL(\fH)$ is written as $\slim$.

\vspace{1mm}
The main objects for our scattering approach are the wave operators which are 
defined as follows.

\bd[Wave operators]
The operators $W_{\d,\gamma}, W_\d, W_\gamma\in\mL(\fH)$, defined by
\ba
\label{Wdgamma}
W_{\d,\gamma}
&:=\slim_{t\to\infty} \e^{-\i t H_\d}\e^{\i tH_\gamma}1_{\rm
ac}(H_\gamma),\\
\label{Wd}
W_\d
&:=\slim_{t\to\infty} \e^{-\i t H_\d}\e^{\i tH},\\
\label{Wgamma}
W_\gamma
&:=\slim_{t\to\infty} \e^{-\i t H}\e^{\i tH_\gamma}1_{\rm
ac}(H_\gamma),
\ea
are called the anisotropy wave operator, the XY wave operator, and the intermediate
wave operator, respectively.
\ed

\br
The Kato-Rosenblum theorem from scattering theory for perturbations of trace class 
type guarantees the existence (and completeness) of \eqref{Wdgamma}, \eqref{Wd}, 
and \eqref{Wgamma} (see, for example, \cite{BW1983} or \cite{Y1998}). 
Indeed, \eqref{vd} and \eqref{v} imply that $V_\d, V\in\mL^0(\fH)$ and, hence, all 
the differences between the corresponding Hamiltonians satisfy 
$H_\d-H_\gamma, H_\d-H, H-H_\gamma\in\mL^1(\fH)$. Moreover, note that 
$1_\ac(H)=1$ since $h$ is the Laplacian on the discrete line (see also the proof of
Proposition \ref{prop:wave} below).
\er

In the next theorem, we determine the 2-point operator of the anisotropy NESS. 

\bt[NESS 2-point operator]
\label{thm:aness}
The anisotropy NESS $\omega_\gamma$ associated with the initial state
$\omega_\d$ and the anisotropy dynamics $\tau^t_\gamma$ exists and its 
2-point operator has the form
\ba
\label{def:Sgam}
S_\gamma
=W_{\d,\gamma}^\ast S_\d W_{\d,\gamma}
+\sum_{e\in\eig(H_\gamma)} 1_e(H_\gamma) S_\d 1_e(H_\gamma).
\ea
\et

\vspace{3mm}

\bprf

\vspace{1mm}
\noindent{\it (a)} {\it Evolution matrix}\\
We start off by studying \eqref{ness} for elements of $\fA$ of the form
$B(F_1)\ldots B(F_{2m})$ for all $m\in\N$ and all $F_1,\ldots, F_{2m}\in\fH$. Since 
the initial state $\omega_\d$ satisfies \eqref{qfs}, we can write
\ba
\label{2mpoint}
\omega_\d(\tau_\gamma^t[B(F_1)\ldots B(F_{2m})])
=\pf([\Omega_{i,j}(t)]_{i,j=1}^{2m}),
\ea
where the following matrix, called the evolution matrix, is defined, for all 
$i,j\in\{1,\ldots, 2m\}$ and all $t\in\R$, by
\ba
\label{evmat}
\Omega_{i,j}(t)
:=(\e^{\i t H_\gamma}\Gamma F_i, S_\d \e^{\i t H_\gamma}F_j),
\ea
and we used the fact from Remark \ref{rem:ham} that 
$[\Gamma,\e^{\i t H_\gamma}]=0$. 

\vspace{1mm}
Next, let us first concentrate on the limit \eqref{ness} for 2-point functions.

\vspace{1mm}
\noindent{\it (b)} {\it Spectral decomposition}\\
Since we know from \cite{HR1986} that $1_\sc(H_\gamma)=0$, 
spectral theory yields the decomposition $1=1_\ac(H_\gamma)+1_\pp(H_\gamma)$ 
which we insert into both arguments of the scalar product \eqref{evmat} between
the propagators and the wave functions. The evolution matrix can thus be written as
$\Omega_{i,j}(t)=\Omega_{i,j}^\ac(t)+\Omega_{i,j}^\pp(t)+R_{i,j}^{(1)}(t)+R_{i,j}^{(2)}(t)$,
where we set
\ba
\label{Omac}
\Omega_{i,j}^\ac(t)
&:=(\e^{\i t H_\gamma}1_\ac(H_\gamma)\Gamma F_i, 
S_\d \e^{\i t H_\gamma}1_\ac(H_\gamma)F_j),\\
\label{Ompp}
\Omega_{i,j}^\pp(t)
&:=(\e^{\i t H_\gamma}1_\pp(H_\gamma)\Gamma F_i, 
S_\d \e^{\i t H_\gamma}1_\pp(H_\gamma)F_j),
\ea
as well as $R_{i,j}^{(1)}(t):=(\e^{\i t H_\gamma}1_\ac(H_\gamma)\Gamma F_i, 
S_\d \e^{\i t H_\gamma}1_\pp(H_\gamma)F_j)$ and analogously for $R_{i,j}^{(2)}(t)$ 
with $1_\ac(H_\gamma)$ and $1_\pp(H_\gamma)$ interchanged.

\vspace{1mm}
\noindent{\it (c)} {\it Large time ergodic mean}\\
In order to study the limit \eqref{ness}, we will treat the foregoing terms separately.

As for the $\ac$-term, using that the initial state is invariant under the decoupled
dynamics, i.e., that $[S_\d, H_\d]=0$, we can write
$
\Omega_{i,j}^\ac(t)
=(\e^{-\i t H_\d}\e^{\i t H_\gamma}1_\ac(H_\gamma)\Gamma F_i, 
S_\d \e^{-\i t H_\d}\e^{\i t H_\gamma}1_\ac(H_\gamma)F_j)
$
and, hence, the large time ergodic mean from \eqref{ness} becomes
\ba
\label{limOmac}
\lim_{t\to\infty}\Omega_{i,j}^\ac(t)
=(\Gamma F_i, W_{\d,\gamma}^\ast S_\d W_{\d,\gamma}F_j),
\ea
and we denote the right hand side of \eqref{limOmac} by $\Omega_{i,j}^\ac$.

Next, we also know from \cite{HR1986} that $\eig(H_\gamma)$ is a finite set 
containing eigenvalues of finite multiplicity only (in the case at hand, the eigenvalues 
of $H_\gamma$ and their corresponding eigenfunctions can be computed explicitly 
but we do not need them to determine the heat flux in Theorem \ref{thm:flux} 
below which is the main purpose of the present study).
Hence, plugging $1_\pp(H_\gamma)=\sum_{e\in\eig(H_\gamma)}1_e(H_\gamma)$
into \eqref{Ompp}, and using that 
$\e^{\i t H_\gamma}1_e(H_\gamma)=\e^{\i t e}1_e(H_\gamma)$, the large time ergodic
mean of the $\pp$-term reads
\ba
\lim_{T\to\infty}\frac{1}{T}\int_0^T\hspace{-1mm}\d t\hspace{1mm}
\Omega_{i,j}^\pp(t)
=\sum_{e\in\eig(H_\gamma)} 
(\Gamma F_i, 1_e(H_\gamma)S_\d 1_e(H_\gamma)F_j).
\ea

Moreover, since $|R_{i,j}^{(1)}(t)|\le \|1_\pp(H_\gamma) S_\d \e^{\i t H_\gamma}
1_\ac(H_\gamma)\Gamma F_i\| \|F_j\|$ due to the Cauchy-Schwarz inequality, and 
since $1_\pp(H_\gamma)\in\mL^0(\fH)$, we know from scattering theory that the 
right hand side of the foregoing estimate vanishes if $t\to\infty$. The term 
$R_{i,j}^{(2)}(t)$ is treated analogously.

\vspace{1mm}
We finally study the large time ergodic mean for observables of the form 
\eqref{2mpoint} and for general $A\in\fA$.

\vspace{1mm}
\noindent{\it (d)} {\it Existence}\\
Since the Pfaffian is a polynomial function of the entries of the matrix on which it 
acts and since $\Omega_{i,j}^\pp(t)$ is a trigonometric polynomial in $t$, the large 
time ergodic mean of $\pf([\Omega_{i,j}^\ac+\Omega_{i,j}^\pp(t)]_{i,j=1}^{2m})$ 
also exists (recall that, in general, the complex-valued functions on $\R$ which are
almost periodic [in the sense of H. Bohr, the brother of N. Bohr] form an algebra with
respect to the usual pointwise linear operations and multiplication, and the large 
time ergodic mean plays the role of a scalar product). Moreover, since, in addition,
$\Omega_{i,j}(t)$ and $\Omega_{i,j}^\ac+\Omega_{i,j}^\pp(t)$ are both uniformly
bounded in $t$, we have $\lim_{t\to\infty} |\pf([\Omega_{i,j}(t)]_{i,j=1}^{2m})
-\pf([\Omega_{i,j}^\ac+\Omega_{i,j}^\pp(t)]_{i,j=1}^{2m})|=0$ which implies that
\ba
\label{meanpoly}
\lim_{T\to\infty}\frac1T\int_0^T\hspace{-1mm}\d t\hspace{1mm}
\omega_\d(\tau^t_\gamma[B(F_1)\ldots B(F_{2m})])
=\lim_{T\to\infty}\frac1T\int_0^T\hspace{-1mm}\d t\hspace{1mm}
\pf([\Omega_{i,j}^\ac+\Omega_{i,j}^\pp(t)]_{i,j=1}^{2m}).
\ea

Finally, since $\fA$ is, by definition, the $C^\ast$\hspace{-0.5mm}-completion 
of the $\ast$-algebra generated by the identity $1$ and the elements $B(F)$ and
$B^\ast(F)$ satisfying \eqref{B1} and \eqref{B2} and since
$|\omega_\d(\tau^t_\gamma(A))|\le \|A\|$ for all $A\in\fA$, the existence of the limits
\eqref{meanpoly} and the uniform convergence in $t$ of the ergodic mean for the
approximant define the anisotropy NESS \eqref{ness}. 

Hence, we arrive at the conclusion.
\eprf

\br
Since we know from \cite{AP2003} that, for the case at hand, the so-called XY 
2-point operator $S:=W^\ast_\d S_\d W_\d$ is given by
\ba
\label{S}
S=(1-s)\oplus\zeta s\zeta,
\ea
where, in momentum space, $s\in\mL(\fh)$ acts through multiplication by the function
$s_{\beta_L,\beta_R}$, where for all $\alpha,\beta\in\R$, we set 
\ba
\label{rhoXY}
s_{\alpha,\beta}(k)
:=
1_{[0, \pi]}(k)\rho_{\alpha}(\cos(k))
+1_{[-\pi, 0]}(k)\rho_{\beta}(\cos(k)),
\ea
and since the chain rule for wave operators from scattering theory implies that
$W_{\d,\gamma}=W_\d W_\gamma$,
we can rewrite the first term on the right hand side of \eqref{def:Sgam} in the form
\ba
\label{WSW}
W_{\d,\gamma}^\ast S_\d W_{\d,\gamma}
=W_\gamma^\ast S W_\gamma.
\ea
\er

\section{Stationary scattering theory}
\label{sec:scattering}

In this section, motivated by \eqref{WSW}, we determine the action of the 
intermediate wave operator $W_\gamma$. To this end, we will make use of the 
so-called stationary approach to scattering theory (see, for example, \cite{BW1983} 
or \cite{Y1998} for a mathematical presentation of the theory and for the ingredients
used in the proof of Proposition \ref{prop:wave} below). 

\vspace{1mm}
In the following, we switch to momentum space,
\ba
\hat\fh
:=L^2\big([-\pi,\pi];\tfrac{{\rm d} k}{2\pi}\big),
\ea
by means of the unitary Fourier transform $\ff:\fh\to\hat\fh$. The latter is defined, 
as usual, by $\ff f:=\sum_{x\in\Z}f(x)\e_x$, where  the plane wave $\e_x\in\hat\fh$ 
is given by $\e_x(k):=\e^{\i k x}$ for all $x\in\Z$. Moreover, we extend it through
$\fF:=\ff\oplus\ff$ to $\widehat\fH:=\hat\fh\oplus\hat\fh$ (the scalar products on
$\hat\fh$ and $\widehat\fH$ are again both denoted by $(\cdot,\cdot)$). For all
$a\in\mL(\fh)$ and all $A\in\mL(\fH)$, we set $\hat a:=\ff a\ff^\ast\in\mL(\hat\fh)$ 
and $\widehat A:=\fF A\fF^\ast\in\mL(\widehat\fH)$.

\bp[Wave operator]
\label{prop:wave}
In momentum space $\widehat\fH$, the action of the wave operator $W_\gamma$ on
completely localized wave functions is given, for all $x\in\Z$, by
\ba
\label{We0}
\widehat W_\gamma\hspace{0.2mm}\e_x\oplus 0
&=\e_x\oplus 0
-\frac{\gamma}{2}
\left(w^{(1)}_{\gamma,x}\oplus 0+
0\oplus w^{(2)}_{\gamma,x}
\right),
\ea
where the functions $w^{(1)}_{\gamma,x}, w^{(2)}_{\gamma,x}\in\hat\fh$ are defined 
by
\ba
w^{(1)}_{\gamma,x}
&:=w^{(1)}_{\gamma,x, a, a+1}+w^{(1)}_{\gamma,x, a+1, a},\\
w^{(2)}_{\gamma,x}
&:=w^{(2)}_{\gamma,x, a, a+1}-w^{(2)}_{\gamma,x, a+1, a},
\ea
and, for all $a_1, a_2\in\Z$, we set
\ba
\label{t1}
w^{(1)}_{\gamma,x,a_1,a_2}(k)
&:=\frac{\gamma}{2}\frac{\e_{a_1}(k)}{D_\gamma(k) \sin^2(k)}
\hspace{0.5mm}\cdot\nonumber\\
&\hspace{4mm}\cdot 
\left[
\e_{|x-a_2|+1}(|k|)+\e_{|x-a_1|}(|k|)
+\i \frac{\gamma^2}{2}\frac{\e_1(|k|)}{\sin(|k|)}
\left(\e_{|x-a_2|+1}(|k|)-\e_{|x-a_1|}(|k|)\right)
\right],\\
w^{(2)}_{\gamma,x,a_1,a_2}(k)
&:=\i (-1)^{x-a_1}\frac{\e_{a_1}(k)}{\overline{D_\gamma(k)}
\sin(|k|)}\hspace{0.5mm}\cdot\nonumber\\
&\hspace{4mm}\cdot \left[
\e_{|x-a_2|}(-|k|)
+\frac{\gamma^2}{2}\frac{\e_1(-|k|)}{\sin^2(k)}
\left(\e_{|x-a_2|}(-|k|)\cos(k)-\e_{|x-a_1|}(-|k|)\right)
\right],
\ea
and also
\ba
\label{Dgam}
D_\gamma(k)
:=1+\gamma^2\frac{\e_1(|k|)}{\sin^2(k)}
\left(\cos(k)-\frac{\gamma^2}{4}\hspace{0.5mm}\e_1(|k|)\right).
\ea
\ep

\vspace{3mm}

\br
Note that
$|D_\gamma(k)|^2\sin^4(k)
=\gamma^4(4-\gamma^2)^2/16
+\gamma^2[(2-\gamma^2)^2+\gamma^2] \sin^2(k)/2
+(1-\gamma^2)^2\sin^4(k)$. Hence, for $|\gamma|=2$, we have, in general, that
$w^{(i)}_{\gamma,x,a_1,a_2}\not\in\hat\fh$ but still $w^{(i)}_{\gamma,x}\in\hat\fh$.
\er

\br
\label{rem:W0e}
Since $\Gamma\delta_x\oplus 0=0\oplus\delta_x$, since $[\Gamma,\e^{-\i tH}]
=[\Gamma,\e^{-\i tH_\gamma}]=0$ due to Remark \ref{rem:ham}, and since 
$[\Gamma,1_{\rm ac}(H_\gamma)]=0$ due to the spectral theorem and the reflection
invariance of the Lebesgue-Borel measure (implying that $\Gamma F$ belongs to the
absolutely continuous subspace of $H_\gamma$ if $F$ does so), we get $[\Gamma,W_\gamma]=0$ and, hence, $\widehat W_\gamma 0\oplus \e_x
=\widehat\Gamma\widehat W_\gamma\e_x\oplus 0$ for all $x\in\Z$.
\er

In the following, for all $a\in\mL(\fh)$ and all $A\in\mL(\fH)$, we denote by 
$r_z(a):=(a-z1)^{-1}$ and $R_z(A):=(A-z1)^{-1}$ the resolvents of $a$ and $A$ at points 
$z$ in the corresponding resolvent sets.

\vspace{3mm}

\bprf

\vspace{1mm}
\noindent{\it (a)} {\it Stationary approach}\\
The stationary approach to scattering theory expresses the wave operator as a weak
abelian limit and subsequently transforms the time dependent propagators into time
independent resolvents with the help of Parseval's identity, i.e., for all $F,G\in\fH$, we have
\ba
\label{weak}
(G, W_\gamma F)
&=\lim_{\varepsilon\to 0^+} 2\varepsilon\int_0^\infty\d t\hspace{1mm}
\e^{-2\varepsilon t} (G, \e^{-\i tH} \e^{\i tH_\gamma} 
1_\ac(H_\gamma) F)\nonumber\\
&=\lim_{\varepsilon\to 0^+} 
\int_{-\infty}^\infty{\rm d}e\hspace{1mm} \frac{\varepsilon}{\pi}
(R_{e-\i\varepsilon}(H)G,
R_{e-\i\varepsilon}(H_\gamma)1_\ac(H_\gamma)F).
\ea

\vspace{0.5mm}
\noindent{\it (b)} {\it Interaction matrix}\\
In order to compute the integrand in \eqref{weak}, we express the resolvent of
$H_\gamma$ in terms of the resolvent of $H$ by means of the 2nd resolvent identity,
 i.e.,  by 
$R_{e-\i\varepsilon}(H_\gamma)
=R_{e-\i\varepsilon}(H)-\gamma R_{e-\i\varepsilon}(H_\gamma) V
R_{e-\i\varepsilon}(H)$. 
For all $e\in\R$, all $\varepsilon>0$, and all $F=f_1\oplus f_2\in\fH$, we then get
\ba
\label{2ndRes}
R_{e-\i\varepsilon}(H_\gamma)F
&=R_{e-\i\varepsilon}(H)F
-\frac{\gamma}{2}R_{e-\i\varepsilon}(H_\gamma )
\Big( 
[r_{-e+\i\varepsilon}(h)f_2](a+1)
\hspace{0.5mm}\delta_a\oplus 0\nonumber\\
&\hspace{55mm}
-[r_{-e+\i\varepsilon}(h)f_2](a)
\hspace{0.5mm}\delta_{a+1}\oplus 0\nonumber\\
&\hspace{55mm}
+[r_{e-\i\varepsilon}(h)f_1](a+1)
\hspace{0.5mm}0\oplus \delta_a\nonumber\\
&\hspace{55mm}
-[r_{e-\i\varepsilon}(h)f_1](a)
\hspace{0.5mm}0\oplus \delta_{a+1}
\Big),
\ea
where we used that 
$R_{e-\i\varepsilon}(H)=r_{e-\i\varepsilon}(h)\oplus [-r_{-e+\i\varepsilon}(h)]$.

Next, we set $E_1:=\delta_a\oplus 0$, $E_2:=\delta_{a+1}\oplus 0$, 
$E_3:=0\oplus \delta_a$, and $E_4:=0\oplus \delta_{a+1}$, take the scalar product of
\eqref{2ndRes} with $G\in\fH$ from the left, and set 
$\mu:=[(G,R_{e-\i\varepsilon}(H_\gamma)\hspace{0.1mm}E_i)]_{i=1}^4\in\C^4$ and
analogously for $\nu\in\C^4$ with $H_\gamma$ replaced by $H$. Then, plugging
successively $F=E_i$ for all $i\in\{1,2,3,4\}$ into the terms from \eqref{2ndRes} and 
defining $\alpha_{e\pm\i\varepsilon}(x):=(\delta_x,r_{e\pm\i\varepsilon}(h)\delta_0)$
for all $e\in\R$, all $\varepsilon>0$, and all $x\in\Z$, we can compactly write the 
resulting four equations as $A_{\gamma, e-\i\varepsilon}\hspace{0.1mm}\mu=\nu$, 
where the interaction matrix $A_{\gamma, e-\i\varepsilon}\in\C^{4\times 4}$ is 
defined by 
\ba
A_{\gamma, e-\i\varepsilon}
:=1
+\frac{\gamma}{2}\big(\alpha_{e-\i\varepsilon}(1)\sigma_1\otimes\sigma_3
+\alpha_{e-\i\varepsilon}(0)\sigma_2\otimes\sigma_2\big),
\ea
and we made use of the symmetries \eqref{hu}, \eqref{htheta}, and \eqref{hxi}, 
implying, in particular, that 
$\alpha_{e-\i\varepsilon}(-x)\linebreak=\alpha_{e-\i\varepsilon}(x)$ and
$\alpha_{-e+\i\varepsilon}(x)=(-1)^{x+1}\alpha_{e-\i\varepsilon}(x)$.

Moreover, since 
$(1+\gamma V R_{e-\i\varepsilon}(H))(1-\gamma V R_{e-\i\varepsilon}(H_\gamma ))=1$
due to the 2nd resolvent identity, since 
$A_{\gamma, e-\i\varepsilon}=[(E_i, (1+\gamma V R_{e-\i\varepsilon}(H))E_j)]_{i,j=1}^4$,
and since the range of $V $ is spanned by $E_1$, $E_2$, $E_3$, and $E_4$,  the 
interaction matrix is invertible. Hence, taking again the scalar product of 
\eqref{2ndRes} with $G\in\fH$ from the left and plugging $\mu$ into the resulting 
equation leads to
\ba
\label{Rgamma}
(G, R_{e-\i\varepsilon}(H_\gamma )F)
=(G,R_{e-\i\varepsilon}(H)F)
-\frac{\gamma}{2}\sum_{i,j=1}^4 
(E_i',R_{e-\i\varepsilon}(H) F)
[A_{\gamma, e-\i\varepsilon}^{-1}]_{i,j}
(G,R_{e-\i\varepsilon}(H) E_j),
\ea
where we set $E_1':=-E_4$, $E_2':=E_3$, $E_3':=E_2$, and $E_4':=-E_1$ (note that, 
using this notation, we can write $V =\sum_{i=1}^4(E_i',\cdot\hspace{0.5mm}) E_i/2$).

\vspace{1mm}
\noindent{\it (c)} {\it Boundary values}\\
We know from stationary scattering theory that, if the limit of 
$\varepsilon(R_{e-\i\varepsilon}(H)G, R_{e-\i\varepsilon}(H_\gamma)F)/\pi$ for
$\varepsilon\to 0^+$ exists for all $F,G\in\fH$ and almost all $e\in\R$ (where here 
and in the following, the set of full measure in $\R$ may depend on $F$ and $G$), 
the limit and the integration in \eqref{weak} can be interchanged, the limit for
$\varepsilon\to 0^+$ of the integrand in \eqref{weak} equals the limit of 
$\varepsilon(R_{e-\i\varepsilon}(H)G, R_{e-\i\varepsilon}(H_\gamma)F)/\pi$ for all
$F, G\in\fH$ and almost all $e\in\R$, and the integral extends over $[-1,1]$ only 
since $\spec(H)=[-1,1]$.

In order to verify the existence in question, we use \eqref{Rgamma} and replace $G$ 
in \eqref{Rgamma} by the term $(\varepsilon/\pi) R_{e-\i\varepsilon}(H)G$. Since
stationary scattering theory also guarantees the existence of the limits 
$\lim_{\varepsilon\to 0^+}(G,R_{e\pm\i\varepsilon}(H)F)$ for almost all $e\in\R$ 
(the existence argument holds for any Hamiltonian), and since 
$(\varepsilon/\pi)R_{e+\i\varepsilon}(H)R_{e-\i\varepsilon}(H)
=(R_{e+\i\varepsilon}(H)-R_{e-\i\varepsilon}(H))/(2\pi\i)$ due to the 1st resolvent
identity, the limits of the 1st, 2nd, and 4th term on the right hand side of 
\eqref{Rgamma} exist. As for the 3rd term, we know from \cite{A2011} that, for all 
$e\in (-1,1)$ and all $x\in\Z$, the limit 
$\alpha_{e-\i 0}(x):=\lim_{\varepsilon\to 0^+}\alpha_{e-\i\varepsilon}(x)$ 
exists and has the form
\ba
\label{aei0}
\alpha_{e-\i 0}(x)
=-\i \frac{\big(e+\i \sqrt{1-e^2}\big)^{|x|}}{\sqrt{1-e^2}}.
\ea
Since the limit from above of the expectation value of the resolvent of $H_\gamma $
(instead of $H$) also exists, and proceeding as in {\it (b)}, we find that the matrix 
$A_{\gamma, e-\i 0}:=\lim_{\varepsilon\to 0^+}A_{\gamma, e-\i\varepsilon}$ is 
invertible, too. Since taking the inverse is a continuous operation, we also have that 
$\lim_{\varepsilon\to 0^+}A_{\gamma, e-\i\varepsilon}^{-1}
=A_{\gamma, e-\i 0}^{-1}$. 

Moreover, for all $e\in (-1,1)$, we can compute the inverse and get
\ba
\label{detAA0}
 A_{\gamma, e-\i 0}^{-1}
=\frac{1}{\det(A_{\gamma, e-\i 0})}\hspace{0.3mm}
\left(\beta^{(1)}_{\gamma,e} 1
+\beta^{(2)}_{\gamma,e}\sigma_3\otimes\sigma_1
+\beta^{(3)}_{\gamma,e}\sigma_1\otimes\sigma_3
+\beta^{(4)}_{\gamma,e}\sigma_2\otimes\sigma_2
\right),
\ea
where we set
\ba
\beta^{(1)}_{\gamma,e} 
&:=1+\frac{\gamma^2}{2}\frac{e\big(e+\i \sqrt{1-e^2}\big)}{1-e^2},\\
\beta^{(2)}_{\gamma,e} 
&:=-\frac{\gamma^2}{2}\frac{e+\i \sqrt{1-e^2}}{1-e^2},\\
\beta^{(3)}_{\gamma,e} 
&:=\i\frac{\gamma}{2}\frac{e+\i \sqrt{1-e^2}}{\sqrt{1-e^2}}
\bigg(1+\i\frac{\gamma^2}{2}\frac{e+\i \sqrt{1-e^2}}{\sqrt{1-e^2}}\bigg),\\\beta^{(4)}_{\gamma,e}
&:=\i\frac{\gamma}{2}\frac{1}{\sqrt{1-e^2}}
\bigg(1-\i\frac{\gamma^2}{2}\frac{e+\i \sqrt{1-e^2}}{\sqrt{1-e^2}}\bigg),
\ea
and the determinant has the form
\ba
\label{detA0}
\det(A_{\gamma, e-\i 0})
=1+\gamma^2\frac{e+\i \sqrt{1-e^2}}{1-e^2}
\bigg(e-\frac{\gamma^2}{4}\big(e+\i \sqrt{1-e^2}\big)\bigg).
\ea

\vspace{1mm}
\noindent{\it (d)} {\it Energy space representation}\\
We next switch to energy space,
\ba
\label{def:L2C2}
\tilde\fh
:=L^2([-1,1],\C^2;\d e),
\ea
by means of the unitary operator $\tilde\ff:\hat\fh\to\tilde\fh$ defined in 
\cite{A2011} on the momentum space $\hat\fh$ as 
$(\tilde\ff\varphi)(e)
:=[\varphi(\arccos(e)), \varphi(-\arccos(e))]/(\sqrt{2\pi} \sqrt[4]{1-e^2})$.  Since we 
want the XY Hamiltonian $H$ to become the operator acting through multiplication 
by the free variable in $\widetilde\fH:=\tilde\fh\oplus\tilde\fh$ (i.e., we want to make 
use of the multiplication operator version of the spectral theorem), we extend 
$\tilde\ff$ to  $\widetilde\fF:=\tilde\ff\oplus\tilde\theta\tilde\xi\hspace{0.3mm}
\tilde\ff$ because of the 2nd factor in $H$, where, for all $a\in\mL(\fh)$ and all
$A\in\mL(\fH)$, we set $\tilde a:=\tilde\ff\hat a\tilde\ff^\ast\in\mL(\tilde\fh)$ and 
$\widetilde A:=\widetilde\fF\widehat A\widetilde\fF^\ast\in\mL(\widetilde\fH)$, and 
we note that $(\tilde\theta\eta)(e)=\sigma_1\eta(e)$ and 
$(\tilde\xi\eta)(e)=\sigma_1\eta(-e)$ for all $\eta\in\tilde\fh$. Hence, since 
$(\hat h\varphi)(k)=\cos(k)\varphi(k)$ for all $\varphi\in\hat\fh$, the XY Hamiltonian 
$H$ indeed becomes the desired multiplication operator under conjugation with 
$\widetilde\fF$ because $(\tilde h\eta)(e)=e \eta(e)$ and 
$(\tilde\theta\tilde\xi \tilde h\tilde\theta\tilde\xi\eta)(e)=-e\eta(e)$ for all 
$\eta\in\tilde\fh$, where we used that 
$(\tilde\theta\tilde\xi)^\ast=\tilde\theta\tilde\xi$ and that, for
all $\eta=[\eta_1,\eta_2]\in\tilde\fh$, the adjoint of $\tilde\ff$ has the form
\ba
\label{tildefstar}
\big({\tilde\ff}^{^\ast}\hspace{-0.5mm}\eta\big)(k)
=\sqrt{2\pi}\sqrt[4]{1-\cos^2(k)}\hspace{0.5mm}
\big[
1_{[0,\pi]}(k)\hspace{0.3mm}\eta_1(\cos(k))
+1_{[-\pi,0]}(k)\hspace{0.3mm}\eta_2(\cos(k))
\big].
\ea

Furthermore, since the spectral core of $H$ equals $\spec(H)=[-1,1]$, stationary 
scattering theory also yields, for all $F=f_1\oplus f_2\in\fH$ and all 
$G=g_1\oplus g_2\in\fH$, that
\ba
\label{fiber}
\lim_{\varepsilon\to 0^+}\frac{\varepsilon}{\pi}
(R_{e-\i\varepsilon}(H)G,R_{e-\i\varepsilon}(H)F)
=\langle \tilde g_1(e), \tilde f_1(e)\rangle 
+\langle (\tilde\theta\tilde\xi\tilde g_2)(e), 
 (\tilde\theta\tilde\xi\tilde f_2)(e)\rangle,
\ea
where we set $\tilde f:=\tilde\ff\ff f$ for all $f\in\fh$, and  
$\langle\hspace{0.3mm}\cdot,\cdot\hspace{0.3mm}\rangle$ stands for the complex
Euclidean scalar product on the constant fiber $\C^2$ of the direct integral 
\eqref{def:L2C2}.

Now, plugging \eqref{Rgamma} into \eqref{weak}, commuting the limit and the
 integration, restricting the integration domain to $[-1,1]$ as discussed in {\it (c)}, 
and substituting \eqref{fiber} into the resulting expression, we get
\ba
\label{Wtilde}
\widetilde W_\gamma\widetilde F
&=\widetilde F
-\frac{\gamma}{2}\sum_{i,j=1}^4 
(E_i',R_{\hspace{0.5mm}\cdot\hspace{0.2mm}-\i0}(H)F)
\hspace{0.2mm}[A_{\gamma, \cdot\hspace{0.2mm}-\i 0}^{-1}]_{i,j}
\hspace{0.5mm}\widetilde E_j,
\ea
where we set $\widetilde F:=\widetilde\fF\fF F$ for all $F\in\fH$
and $(E_i',R_{e-\i 0}(H)F):=\lim_{\varepsilon\to
0^+}(E_i',R_{e-\i\varepsilon}(H)F)$ exists as discussed in {\it (c)}.

\vspace{1mm}
\noindent{\it (e)} {\it Momentum space representation}\\
Next, let $x\in\Z$ and plug $F=\delta_x\oplus 0$ into \eqref{Wtilde}. Then, applying 
$\widetilde\fF=\tilde\ff\oplus\tilde\theta\tilde\xi\hspace{0.3mm}\tilde\ff$ on both
sides of \eqref{Wtilde} and noting that 
$(E_1',R_{\hspace{0.5mm}\cdot\hspace{0.2mm}-\i0}(H)\delta_x\oplus 0)
=(E_2',R_{\hspace{0.5mm}\cdot\hspace{0.2mm}-\i0}(H)\delta_x\oplus 0)
=0$ 
and that
\ba
\label{E3'}
(E_3',R_{\hspace{0.5mm}\cdot\hspace{0.2mm}-\i0}(H)\delta_x\oplus 0)
&=\alpha_{\hspace{0.5mm}\cdot\hspace{0.2mm}-\i0}(x-(a+1)),\\
\label{E4'}
(E_4',R_{\hspace{0.5mm}\cdot\hspace{0.2mm}-\i0}(H)\delta_x\oplus 0)
&=-\alpha_{\hspace{0.5mm}\cdot\hspace{0.2mm}-\i0}(x-a),
\ea
the action,  in momentum space $\widehat\fH$, of the intermediate wave operator
$W_\gamma$ on completely localized wave functions becomes
\ba
\label{whatW}
\widehat W_\gamma\e_x\oplus 0
&=\e_x\oplus 0\nonumber\\
&-\frac{\gamma}{2}
\sum_{j=1}^4
{\tilde\ff}^{^\ast}\hspace{-0.6mm}\oplus{\tilde\ff}^{^\ast}
\hspace{-0.2mm}\tilde\theta\tilde\xi
\Big[\big(
\alpha_{\hspace{0.5mm}\cdot\hspace{0.2mm}-\i0}(x-(a+1))
\hspace{0.2mm}[A_{\gamma, \cdot\hspace{0.2mm}-\i 0}^{-1}]_{3,j}
-\alpha_{\hspace{0.5mm}\cdot\hspace{0.2mm}-\i0}(x-a)
\hspace{0.2mm}[A_{\gamma, \cdot\hspace{0.2mm}-\i 0}^{-1}]_{4,j}\big)
\hspace{0.5mm}\widetilde E_j
\Big].
\ea
Finally, we plug \eqref{aei0} -- \eqref{detA0} and \eqref{E3'} -- \eqref{E4'} into \eqref{whatW} and use \eqref{tildefstar} and the fact that $\tilde\delta_x(e)
=[(e+\i\sqrt{1-e^2})^x, (e-\i\sqrt{1-e^2})^x]/(\sqrt{2\pi}\sqrt[4]{1-e^2})$ for all $x\in\Z$.

Hence, we arrive at the conclusion.
\eprf

\section{Entropy production rate}
\label{sec:entropy}

In this section, we determine the expectation value in the anisotropy NESS of the
energy flowing between the reservoirs through the sample system.

\vspace{1mm}
In the following, we will make use of the second quantization $b$ in the selfdual 
framework introduced in Definition \ref{def:obs}{\it (d)}.

\bd[Entropy production]
\label{def:heat}
\hspace{0mm}

\vspace{1mm}
\noindent{\it (a)} 1-particle energy current\\
The energy flow from the left reservoir into the sample is described by the 
1-particle energy current observable $\Phi\in\mL^0(\fH)$ given by
\ba
\label{def:Phi}
\Phi
:=-\left.\frac{\rm d}{{\rm d}t}\right|_{t=0}
\e^{\i t H_\gamma } H_L\e^{-\i t H_\gamma}.
\ea

\vspace{1mm}
\noindent{\it (b)} Heat flux\\
The heat flux $J_\gamma$ is defined to be the NESS expectation value of the 
extensive energy current observable, i.e., 
\ba
\label{def:flux}
J_\gamma
:=\omega_\gamma(b(\Phi)).
\ea
Moreover, the entropy production rate is given by 
$\sigma_\gamma:=(\beta_R-\beta_L) J_\gamma$.

\ed

\br
\label{rem:1stlaw}
Let us denote by $J_{\gamma,R}$ the NESS expectation value of the extensive energy
current observable $b(\Phi_R)$ whose 1-particle observable $\Phi_R$ describes the
energy flow from the right reservoir into the sample, i.e., $\Phi_R$ is defined as in
\eqref{def:Phi} but with $H_L$ replaced by $H_R$. We thus get that the sum of the
derivative (in the Banach space $\mL(\fH)$) from \eqref{def:Phi} and its analog for 
the right reservoir can be written as $\Phi+\Phi_R=-\i [H_\gamma, H_R+H_L]
=\i [H_\gamma, Q]$, where we set $Q:=H_\gamma-(H_R+H_R)$, and we note that
$Q\in\mL^0(\fH)$ because $Q=H_S+V_\d+\gamma V$. Here, as in Definition 
\ref{def:initial}{\it (b)},  $h_S\in\mL(\fh)$ is the 1-particle sample Hamiltonian defined 
by $h_S:=p_Shp_S$, and the orthogonal projection $p_S\in\mL(\fh)$ is given by
$p_Sf:=1_{\Z_S}f$ for all $f\in\fh$. Moreover, the lifting of $h_S$ to $\mL(\fH)$ is
denoted by $H_S:=h_S\sigma_3$. 

Since \eqref{sdsq} and \eqref{2ptf} imply that 
$\omega_\gamma(b(A))=-\tr(S_\gamma A)$ for all $A\in\mL^0(\fH)$ with 
$A=A^\ast$ and $\Gamma A\Gamma=-A$, where $\tr(\cdot)$ stands for the trace on
$\mL^1(\fH)$, we get, with Remark \ref{rem:ham}, that
\ba
\label{1stlaw}
J_\gamma+J_{\gamma,R}
=-\i\hspace{0.5mm} \tr(S_\gamma [H_\gamma, Q]).
\ea
Since it follows from \eqref{ness} that the anisotropy NESS $\omega_\gamma$ is
invariant with respect to the anisotropy dynamics, i.e., since
$\omega_\gamma(\tau_\gamma^t(A))=\omega_\gamma(A)$ 
for all $t\in\R$ and all $A\in\fA$, we have $[S_\gamma,H_\gamma]=0$. Therefore, due 
to the cyclicity of the trace (i.e., $\tr(AB)=\tr(BA)$ for all $A\in\mL^1(\fH)$ and all
$B\in\mL(\fH)$), \eqref{1stlaw} implies that $J_\gamma+J_{\gamma,R}=0$,
i.e., we obtain the first law of thermodynamics.
\er

\br
Due to Remark \ref{rem:1stlaw}, the entropy production from Definition 
\ref{def:heat}{\it (b)} has its usual form, i.e., we can write 
$\sigma_\gamma=-(\beta_R J_{\gamma, R}+\beta_L J_\gamma)$.
\er

\vspace{5mm}

\begin{figure}
\centering
\begin{tikzpicture}
\node (flux) {\includegraphics[width=75mm,height=50mm]{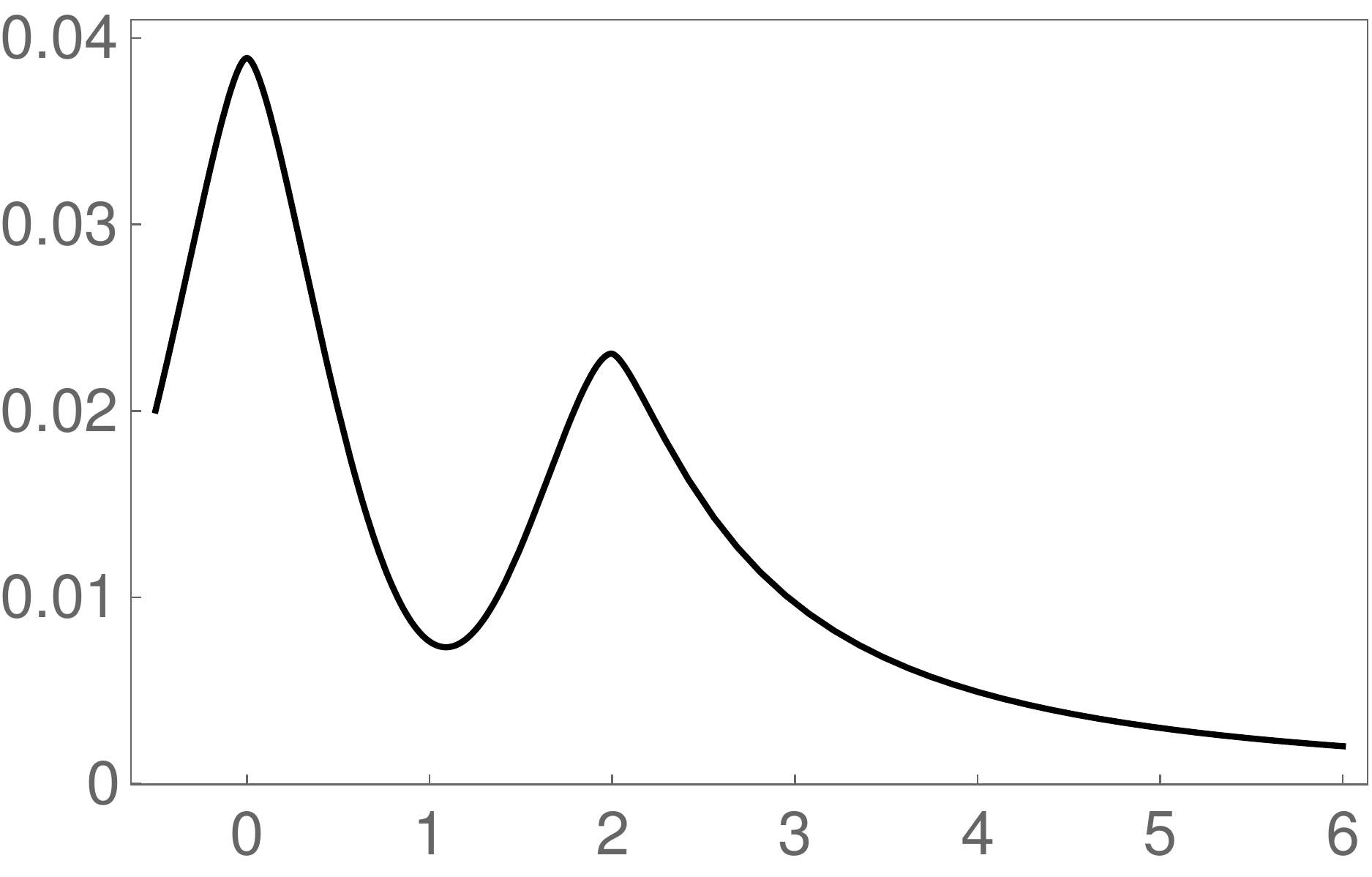}};
\node[below=of flux, anchor=center, xshift=3.5mm, yshift=8mm] 
{Anisotropy $\gamma$};
\node[left=of flux, rotate=90, anchor=center, xshift=2mm, yshift=-6.5mm] 
{Heat flux $J_\gamma$};
\end{tikzpicture}
\vspace{-3.5mm}
\caption{The heat flux $\gamma\mapsto J_\gamma$ 
for $\beta_L=1$ and $\beta_R=2$ (being an even function of the strength $\gamma$
of the anisotropy perturbation). }
\label{fig:flux}
\end{figure}

We now arrive at the main result of our study (see Figure
\ref{fig:flux}).

\bt[Heat flux]
\label{thm:flux}
For all $\gamma\in\R$, the heat flux has the form
\ba
\label{flux}
J_\gamma
=\frac12 \intg\hspace{0.5mm}
\sin(2|k|)\hspace{0.1mm}\Delta(\cos(k))
\left[1-\frac{P_\gamma(\sin(k))}{Q_\gamma(\sin(k))}\right],
\ea
where we set $\Delta:=\rho_{\beta_L}-\rho_{\beta_R}$ and the even polynomials
\ba
\label{PgamQgam}
P_\gamma(e)
&:=\frac{b_\gamma}{2}\hspace{0.2mm} e^2+c_\gamma,\\
Q_\gamma(e)
&:=a_\gamma e^4+b_\gamma e^2+c_\gamma,
\ea
have the nonnegative coefficients 
\ba
\label{agam}
a_\gamma
&:=\left(1-\gamma^2\right)^2,\\
\label{bgam}
b_\gamma
&:=\frac{\gamma^2}{2}\left[\left(2-\gamma^2\right)^2+\gamma^2\right],\\
\label{cgam}
c_\gamma
&:=\frac{\gamma^4}{16}\left(4-\gamma^2\right)^2.
\ea
\et

\vspace{4mm}

\br
\label{rem:FermiDiff}
Since $1/(1+\e^x)-1/(1+\e^y)=\sinh([y-x]/2)/(\cosh([y-x]/2)+\cosh([y+x]/2))$ for all
$x,y\in\R$, the difference of the Fermi-Dirac functions can be written as
\ba
\label{DFermi}
\Delta(e)
=\frac{\sinh([\beta_R-\beta_L]e/2)}
{\cosh([\beta_R-\beta_L]e/2)+\cosh([\beta_R+\beta_L]e/2)}.
\ea
\er

\br
Note also that \eqref{flux} is independent of the sample size $n$ and of the
supporting sites $\{a, a+1\}$ of the local anisotropy perturbation.
\er

\br
In the light of \eqref{XYDensity}, the values $\gamma=1$ and $\gamma=2$ correspond
to a vanishing and, up to a global sign, an isotropic $\sigma_2$-contribution, 
respectively (see Figure \ref{fig:flux}). Moreover, Corollary \ref{cor:ep} below yields 
that $J_\gamma>0$. The regularity of the heat flux function 
$\gamma\mapsto J_\gamma$ will be discussed elsewhere (see also \cite{A2016}).
\er

\vspace{3mm}

\bprf

\vspace{1mm}
\noindent{\it (a)}  {\it Flux structure}\\
The 1-particle energy current observable \eqref{def:Phi} has the form
\ba
\label{Phi}
\Phi
&=-\i [H_\gamma, H_L]\nonumber\\
&=-\i[h,h_L]\sigma_0+\gamma\{v,h_L\}\sigma_1\nonumber\\
&=\frac{1}{2}\hspace{0.5mm}\Im[p_{-(n+2),-n}]\sigma_0,
\ea
where, in the last equality, we used that $vh_{\rm L}=h_{\rm L}v =0$ due to \eqref{a}.

Since $\Phi=\Phi^\ast\in\mL^0(\fH)$ and $\Gamma\Phi\Gamma=-\Phi$, we have
$\omega_\gamma(b(\Phi))=-\tr(S_\gamma \Phi)$ as in Remark \ref{rem:1stlaw}. Hence,
using \eqref{def:Sgam} and \eqref{WSW}, we write $J_{\gamma}
=-(J_{\gamma, \ac}+J_{\gamma, \pp})$ and we define
\ba
\label{Jac}
J_{\gamma, \ac}
&:=\tr(W_\gamma^\ast S W_\gamma \Phi),\\
\label{Jpp}
J_{\gamma, \pp}
&:=\sum_{e\in\eig(H_\gamma)}\tr(1_e(H_\gamma) S_\d1_e(H_\gamma)\Phi).
\ea

\vspace{1mm}
\noindent{\it (b)} {\it Term $J_{\gamma, \ac}$}\\
Using \eqref{Phi} and computing the trace (for example with respect to the 
orthonormal basis $\{\delta_x\oplus 0, 0\oplus \delta_x\}_{x\in\Z}$ of $\fH$), we get
$J_{\gamma, \ac}=J_{\gamma, \ac}^{(1)}+J_{\gamma, \ac}^{(2)}$, where we define
\ba
\label{Jac1}
J_{\gamma, \ac}^{(1)}
&:=\frac12\hspace{0.5mm}
\Im[(W_\gamma\delta_{-(n+2)}\oplus 0,  
S W_\gamma\delta_{-n}\oplus 0)],\\
\label{Jac2}
J_{\gamma, \ac}^{(2)}
&:=\frac12\hspace{0.5mm}
\Im[(W_\gamma 0\oplus\delta_{-(n+2)}, 
S W_\gamma 0\oplus\delta_{-n})].
\ea

Let us first determine \eqref{Jac1} by switching to momentum space and by specializing
\eqref{We0} -- \eqref{Dgam} to the case $x=-n$ and $x=-(n+2)$. Using \eqref{a}, we 
can write
\ba
\label{w1aa+1}
w^{(1)}_{\gamma,-n,a,a+1}(k)
&=\gamma\hspace{0.5mm} 
\frac{\e_{a}(|k|+k)\e_{n+1}(|k|)}{D_\gamma(k)\sin^2(k)}
\left(\cos(k)-\frac{\gamma^2}{2}\e_{1}(|k|)\right),\\
w^{(1)}_{\gamma,-n,a+1,a}(k)
&=\gamma\hspace{0.5mm}\frac{\e_{a+1}(|k|+k)\e_{n}(|k|)}
{D_\gamma(k)\sin^2(k)},
\ea
and also
\ba
w^{(2)}_{\gamma,-n,a,a+1}(k)
&=\i (-1)^{n+a} 
\frac{\e_{a}(k-|k|)\e_{n+1}(-|k|)}{\overline{D_\gamma(k)}\sin^2(k)}
\left(\sin(|k|)-\i \frac{\gamma^2}{2}\e_{1}(-|k|)\right),\\
\label{w2a+1a}
w^{(2)}_{\gamma,-n,a+1,a}(k)
&=\i (-1)^{n+a} 
\frac{\e_{a}(k-|k|)\e_{n}(-|k|)\e_{1}(k)}
{\overline{D_\gamma(k)}\sin^2(k)}
\left(\sin(|k|)+\i \frac{\gamma^2}{2}\e_{1}(-|k|)\right).
\ea
Moreover, for $(a_1,a_2)\in\{(a,a+1), (a+1,a)\}$, we get 
$w^{(1)}_{\gamma,-(n+2),a_1,a_2}(k)=\e_2(|k|)w^{(1)}_{\gamma,-n,a_1,a_2}(k)$ and 
$w^{(2)}_{\gamma,-(n+2),a_1,a_2}(k)=\e_2(-|k|)w^{(2)}_{\gamma,-n,a_1,a_2}(k)$. Next, 
we plug \eqref{w1aa+1} -- \eqref{w2a+1a} and \eqref{S} into the scalar product 
on the right hand side of \eqref{Jac1} and note that $\ff (1-s)\ff^\ast$ and 
$\ff \zeta s\zeta\ff^\ast$ act through multiplication by the functions 
$s_{-\beta_L, -\beta_R}$ and $s_{{\beta_R}, {\beta_L}}$ from \eqref{rhoXY}, respectively.
Decomposing the resulting expressions with respect to positive and negative momenta,
taking the imaginary parts, regrouping with respect to the inverse temperatures, and
using that, for all $\alpha, \beta\in\R$, we have 
$\rho_{-\alpha}-\rho_{-\beta}=-(\rho_{\alpha}-\rho_{\beta})$ due to Remark 
\ref{rem:FermiDiff} and $\rho_{-\alpha}(\cos(k))\mapsto\rho_{\alpha}(\cos(k))$ if
$k\mapsto k+\pi$ (the symmetry $\xi$ from Remark \ref{rem:sym}), a lengthy
computation yields minus the right hand side of \eqref{flux} divided by $2$.

We next turn to \eqref{Jac2}. Using Remarks \ref{rem:qfs} and \ref{rem:W0e}, we can
write
\ba
\label{Jac2-1}
J_{\gamma, \ac}^{(2)}
=J_{\gamma, \ac}^{(1)}
-\frac12\hspace{0.5mm}
\Im[(W_\gamma\delta_{-(n+2)}\oplus 0, W_\gamma\delta_{-n}\oplus 0)].
\ea
Since, for all $(\alpha,\beta)\in\{(-\beta_L,-\beta_R), (\beta_R,\beta_L)\}$, we have
$\lim_{\beta_R\to 0}\lim_{\beta_L\to 0} s_{\alpha,\beta}(k)=1/2$ and 
$|s_{\alpha,\beta}(k)|\le 1$ for all $k\in [-\pi,\pi]$, Lebesgue's dominated convergence
theorem implies, on one hand, that $\lim_{\beta_R\to 0}\lim_{\beta_L\to 0} 
J_{\gamma, \ac}^{(1)}$ is equal to minus the second term on the right hand 
side of \eqref{Jac2-1} divided by $2$. On the other hand, it also implies that
$\lim_{\beta_R\to 0}\lim_{\beta_L\to 0} J_{\gamma, \ac}^{(1)}=0$ since we know 
from above that $J_{\gamma, \ac}^{(1)}$ is equal to minus the right hand side 
of \eqref{flux} divided by $2$, since 
$\lim_{\beta_R\to 0}\lim_{\beta_L\to 0}\Delta(\cos(k))=0$ for all $k\in [-\pi,\pi]$, and 
since $|\Delta(\cos(k))|\le 1/2$ due to \eqref{DFermi} and  
$0\le 1-P_\gamma(\sin(k))/Q_\gamma(\sin(k))\le 1$  for all $\gamma\in\R$ and all 
$k\in [-\pi,\pi]$ due to \eqref{PgamQgam} -- \eqref{cgam}.
Hence, we find that $J_{\gamma, \ac}=2J_{\gamma, \ac}^{(1)}$.

\vspace{1mm}
\noindent{\it (c)} {\it Term $J_{\gamma, \pp}$}\\
Using the cyclicity of the trace, all the summands on the right hand side of \eqref{Jpp} 
can be written as $\tr(S_\d 1_e(H_\gamma)\Phi 1_e(H_\gamma))$. Since 
$1_e(H_\gamma)\Phi 1_e(H_\gamma)=-\i 1_e(H_\gamma)[H_\gamma, H_L]1_e(H_\gamma)
=-\i 1_e(H_\gamma)(e H_L-H_L e)1_e(H_\gamma)=0$, we get $J_{\gamma, \pp}=0$.

Hence, we arrive at the conclusion.
\eprf

\br
Since scattering theory yields that $W_\gamma^\ast W_\gamma=1_\ac(H_\gamma)$, 
and since $1_\ac(H_\gamma)=1-1_\pp(H_\gamma)$, the second term on the right hand
side of \eqref{Jac2-1} can also be determined using the eigenfunctions of $H_\gamma$.
\er

Finally, we derive the strict positivity of the entropy production for the case at hand, 
i.e., in particular, we obtain the second law of thermodynamics. 

\vspace{3mm}

\bc[Entropy production]
\label{cor:ep}
For all $\gamma\in\R$, we have $0<\sigma_\gamma\le (\beta_R-\beta_L)/2$.
\ec

\bprf

\vspace{1mm}
\noindent{\it (a)} {\it Case $\gamma\neq 0$}\\
Due to the symmetry properties of the integrand in \eqref{flux}, the flux can be 
rewritten as an integral over the domain $[0,\pi/2]$ on which all the factors of the 
integrand are nonnegative. Using, on this domain, the straightforward estimates 
$1-P_\gamma(\sin(k))/Q_\gamma(\sin(k))
\ge b_\gamma \sin^2(k)/[2(a_\gamma+b_\gamma+c_\gamma)]$ and 
$\Delta(\cos(k))\ge \sinh(\delta \cos(k))/e_0$, where we set 
$e_0:=\cosh(\delta)+\cosh(\beta)$ with $\delta:=(\beta_R-\beta_L)/2$ and
$\beta:=(\beta_R+\beta_L)/2$, and carrying out the remaining integration
$d_0:=\int_0^{\pi/2}{\rm d}k/(4\pi)\sin(2k)\sin^2(k)\sinh(\delta \cos(k))
=[(3+\delta^2)\sinh(\delta)-3\delta\cosh(\delta)]/(\pi\delta^4)$, we get the lower 
bound $J_\gamma\ge 2b_\gamma d_0/[(a_\gamma+b_\gamma+c_\gamma)e_0]>0$.

\vspace{1mm}
\noindent{\it (b)} {\it Case $\gamma=0$}\\
Since $P_\gamma(\sin(k))=0$ due to \eqref{bgam} and \eqref{cgam}, we get the lower
bound $J_\gamma\ge 4 d_1/e_0>0$ as in {\it (a)}, where 
$d_1:=\int_0^{\pi/2}{\rm d}k/(4\pi)\sin(2k)\sinh(\delta \cos(k))
=[\delta\cosh(\delta)-\sinh(\delta)]/(2\pi\delta^2)$ (we already know from 
\cite{AP2003} that $J_\gamma>0$ in this case).

\vspace{1mm}
Hence, as soon as the system is truly out of equilibrium, i.e., if $\beta_R>\beta_L$, 
there exists a nonvanishing heat flux flowing through the sample from the left (hotter)
 to the right (colder) reservoir. Due to \eqref{betas} and Definition 
 \ref{def:heat}{\it (b)}, we also find that $\sigma_\gamma >0$. 
Moreover, the upper bound follows from the estimates after \eqref{Jac2-1} in part 
{\it (b)} of the proof of Theorem \ref{thm:flux}.

Hence, we arrive at the conclusion.
\eprf

\br
The upper bound can also be derived directly since $|\omega_\gamma(b(\Phi))|
\le \|b(\Phi)\|=\|\Phi\|_1/2$, where $\|\cdot\|_1$ stands for the trace norm, and we 
used \cite{Araki1987} for the last equality.
\er



\begin{thebibliography}{10} 

\bibitem{Araki1987} Araki H 1987 
{\it Bogoliubov automorphisms and Fock representations of canonical 
anticommutation relations}
Contemp. Math. 62 23

\bibitem{Araki1984} Araki H 1984 
{\it On the XY-model on two-sided infinite chain}
Publ. RIMS Kyoto Univ. 20 277

\bibitem{Araki1971} Araki H 1971
{\it On quasifree states of CAR and Bogoliubov automorphisms}
Publ. RIMS Kyoto Univ. 6 385

\bibitem{A2016} Aschbacher W H 2016
{\it On a quantum phase transition in a steady state out of equilibrium}
J. Phys. A: Math. Theor. 49 415201

\bibitem{A2011} Aschbacher W H 2011
{\it Broken translation invariance in quasifree fermionic correlations out of 
equilibrium}
J. Funct. Anal. 260 3429

\bibitem{AP2003} Aschbacher W H and Pillet C-A 2003
{\it Non-equilibrium steady states of the XY chain} 
J. Stat. Phys. 112 1153

\bibitem{BW1983} Baumg\"artel H and Wollenberg M 1983
{\it Mathematical scattering theory} 
(Birkh\"auser)

\bibitem{BR} Bratteli O and Robinson D W 1987/1997
{\it Operator algebras and quantum statistical mechanics 1/2}
(Springer) 

\bibitem{CSP1969} Culvahouse J W, Schinke D P, and Pfortmiller L G 1969
{\it Spin-spin interaction constants from the hyperfine structure of coupled ions}
Phys. Rev. 177 454 

\bibitem{HR1986} Hume L and Robinson D W 1986
{\it Return to equilibrium in the XY model}
J. Stat. Phys. 44 829 

\bibitem{LSM1961}Lieb E, Schultz T, and Mattis D 1961 
{\it Two soluble models of an antiferromagnetic chain}
Ann. Physics 16 407 

\bibitem{MK2004} Mikeska H-J and Kolezhuk A K 2004
{\it One-dimensional magnetism} in Schollw\"ock U, Richter J, Farnell D J J, and 
Bishop R F (Ed.)
{\it Quantum Magnetism} Lect. Notes Phys. 645 1
(Springer)

\bibitem{R2001} Ruelle D 2001
{\it Entropy production in quantum spin systems}
Commun. Math. Phys. 224 3

\bibitem{Y1998} Yafaev D R 1998
{\it Mathematical scattering theory: general theory}
(AMS)

\end{thebibliography}
\end{document}